\documentclass[preprint]{elsarticle}

\usepackage{lineno,hyperref}
\usepackage{color}
\usepackage{subfigure}
\usepackage{amssymb,amsmath}
\usepackage{indentfirst}

\usepackage{bm}
\usepackage{float}
\usepackage{graphicx}
\usepackage{multirow}
\usepackage{tabularx}
\usepackage{soul}

\usepackage{xcolor}
\usepackage{colortbl,booktabs}
\usepackage{threeparttable}

\soulregister\cite7
\soulregister\citep7
\soulregister\citet7
\soulregister\ref7
\soulregister\pageref7

\modulolinenumbers[5]

\journal{Journal of \LaTeX\ Templates}

\begin{document}

\begin{frontmatter}

\title{Modified QPSK Partition Algorithm Based on MAP Estimation for Probabilistically-Shaped 16-QAM}

\cortext[cor1]{Corresponding author}
\author[a]{Jin Hu}
\author[a]{Zhongliang Sun}
\author[a]{Xuekai Xu}
\author[a]{Mengqi Guo}
\author[a]{Xizi Tang}
\author[focal]{Yueming Lu}
\author[a]{Yaojun Qiao\corref{cor1}}
\ead{qiao@bupt.edu.cn}

\address[a]{State Key Laboratory of Information Photonics and Optical Communications, School of Information and Communication Engineering, Beijing University of Posts and Telecommunications, Beijing 100876, China}
\address[focal]{Key Laboratory of Trustworthy Distributed Computing and Service, Ministry of Education, School of Cyberspace Security, Beijing University of Posts and Telecommunications, Beijing 100876, China}

\begin{abstract}

Probabilistic shaping (PS) is investigated as a potential technique to approach the Shannon limit. However, it has been proved that conventional carrier phase recovery (CPR) algorithm designed for uniform distribution may have extra penalty in PS systems. In this paper, we find that the performance of QPSK partition algorithm is degenerated when PS is implemented. To solve this issue, a modified QPSK partition algorithm that jointly optimizes the amplitude decision threshold and filter weight is proposed, where the optimization of decision threshold is based on maximum $a$ $posterior$ probability (MAP) estimation. Different from the conventional decision methods which commonly use Euclidean distance metric, the MAP-based decision introduces the statistical characteristics of the received signals to obtain an accurate amplitude partition. In addition, the filter weight is optimized for different decision thresholds to enhance the tolerance of ASE-induced phase noise. We verify the feasibility of the proposed algorithm in a 56 GBaud PS 16-ary quadrature amplitude modulation (16-QAM) system. The proposed algorithm reduces the error of phase noise estimation by nearly half. Compared with conventional QPSK partition, the proposed algorithm could narrow the gap with theoretical mutual information (MI) by more than 0.1 bit/symbol. The channel capacity is increased by 4.2\%, 4.3\% and 3.6\% with signal-to-noise ratio (SNR) from 8 dB to 10 dB respectively. These observations show that the proposed algorithm is a promising method to relieve the penalty of QPSK partition algorithm in PS systems.

\end{abstract}


\end{frontmatter}

\section{Introduction}

With the emergence of big data, cloud computing and various intelligent applications, the demand for data transmission rate has shown an explosive growth trend in the past few years. With this overwhelming trend, the optical fiber communication system, which carries 99\% of the network data traffic, needs to continuously increase the transmission capacity to provide higher quality services \cite{bayvel2016maximizing}. Recently, constellation shaping (CS) has been widely investigated as a viable alternative to improve the capacity in optical fiber communication systems \cite{schaedler2020neural,nakamura2020entropy,matsushita202041,karout2017achievable}, by approximating the Gaussian distribution source to close in the Shannon limit \cite{shannon1948mathematical}. CS could be divided into geometric shaping (GS) and probabilistic shaping (PS). Different to the regular QAM formats, GS modifies the symbol locations to set a non-equidistant constellation distribution \cite{chen2018increasing,qu2019probabilistic}. In comparison, PS sends the constellation points with different probabilities \cite{bocherer2015bandwidth,cho2019probabilistic}. In practical, PS is widely applied since it is easier to match the channel conditions and offers more flexible information rate which could be used for rate adaption. Moreover, with the increasing demand for data rate, 16-QAM has been studied extensively as a potential solution for next-generation high-speed optical transmission. The 16-QAM modulation can relieve the hardware requirement, which offers a cost-efficient solution \cite{qu2019probabilistic}. The latest works have shown great interests in the combination of 16-QAM and PS to realize high-speed transmission \cite{loussouarn2020probabilistic,wang2019transmission,zhang2019real,kong2019wdm,fallahpour202016}, which indicate the potential of PS 16-QAM for 200G upgrade transmission.

To investigate the advantages of PS for coherent systems, the researches on digital signal process (DSP) are indispensable. However, the conventional DSP algorithms designed for the uniform distribution might have extra penalty in PS systems. In this regard, many works have been done to investigate the impact of PS on clock recovery, equalization and frequency offset recovery \cite{barbosa2020clock,dris2019blind,yan2019blind,barbosa2019phase}, all showing significant impairment with the common algorithms designed for regular QAM signals. As an important part of DSP, carrier phase recovery (CPR) is also an attractive topic. Recently, the tolerance of laser phase noise using pilot-aided digital phase locked loop (PLL) for uniform distribution and probabilistic shaping is compared, and the further research on the pilot ratio has been evaluated experimentally \cite{okamoto2018laser,sasai2019experimental}. Moreover, the impact of PS on the performance of blind phase search (BPS) algorithm and the cycle-slip probability is also studied in depth \cite{barbosa2018impact,mello2018interplay}. The results show that the BPS performance and cycle-slip probability are strongly affected by the strength of shaping, and the MI gain of PS is degraded or even turned into a great penalty after BPS. However, the QPSK partition algorithm designed for 16-QAM \cite{seimetz2006performance}, which has advantages of low complexity and acceptable linewidth tolerance, has been little studied in PS systems.

In this paper, we find that, similar to the BPS, the QPSK partition algorithm also suffers from MI impairment in PS systems. This is mainly because PS reduces the probability of outer constellation points, which have a more outstanding performance in CPR. Furthermore, the conventional amplitude decision threshold in QPSK partition algorithm is determined by Euclidean distance, which needs to be optimized in PS system considering the a-priori probability. We present a modified QPSK partition algorithm jointly optimizing the decision threshold and the filter weight of different amplitude levels. The optimal decision threshold is obtained to minimize the probability of wrong decisions through the MAP estimation. The weight of outer constellation points is optimized when ASE noise is filtered. The proposed algorithm is verified by numerical simulation in a 56 GBaud PS-16QAM system. The simulation results show that giving more weight to outer constellation points yields a better estimation. By exploiting the proposed optimization of decision threshold and filter weight, the estimation error of CPR is reduced to nearly half. The PS-induced MI penalty can be relieved by more than 0.1 bit/symbol, which could be understood as an increase in channel capacity.

\section{Principle}

\subsection{Signal Model}

Probabilistic shaping optimizes the channel capacity by changing the a-priori probability with the Maxwell-Boltzmann distribution in AWGN channel \cite{kschischang1993optimal}. The probability mass function (PMF) is given by
\begin{eqnarray}
P_{X}(x_{i})=\frac{1}{\sum_{j=1}^{M}{\rm exp}(-\lambda|x_j|^{2})}{\rm exp}(-\lambda|x_i|^{2}).
\end{eqnarray}
The variable $X$ represents the channel input with realizations $ x_1,x_2,\ldots,x_M $, $M$ is the constellation size. $\lambda$ denotes the shaping factor, representing the strength of the shaping. $\lambda$ varies from 0 to 1, and $\lambda=0$ denotes the uniform distribution. The theoretical optimal value of $\lambda$ is determined by the modulation format and SNR.

Assuming that channel impairments and frequency offset have been fully compensated before CPR, the $k_{\rm th}$ received symbol $y_k$ can be expressed as
\begin{eqnarray}
y_k=x_k{\rm exp}(j\varphi_k)+n_k.
\end{eqnarray}
where $x_k$ is the $k_{\rm th}$ transmitted symbol, $\varphi_k$ denotes the laser phase noise and $n_k$ represents the Gaussian noise with variance $\sigma_n^{2}$. Laser phase noise $\varphi_k$ could be modeled as a Wiener process \cite{pfau2009hardware}:
\begin{eqnarray}
\varphi_k=\varphi_{k-1}+w_k.
\end{eqnarray}
$w_k$ is a Gaussian-distributed random variables with zero mean and the variance could be represented as $\sigma_p^2=2\pi\Delta f \tau$. Parameter $\Delta f$ is given by the sum of linewidth of the transmitter and local oscillator laser, and $\tau$ is symbol duration. The purpose of CPR is to estimate the phase noise $\varphi_k$.

In this paper, we use the MI as the evaluation criterion. MI is an upper bound on the achievable information rate, defined as
\begin{eqnarray}
I(X,Y)=H(X)-H(X|Y).
\end{eqnarray}
where $X$ is the transmitted symbols and $Y$ is the received symbols, $H(X)$ is the entropy of $X$, and $H(X|Y)$ represents the reduction of the information due to various interference.

\subsection{Principle of Modified QPSK Partition Algorithm}

\begin{figure}[!t]
\centering
\includegraphics[height= 3.8cm] {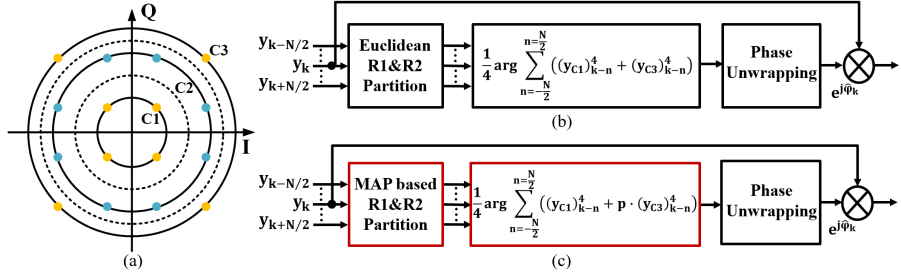}
\caption{(a) Partitioning of 16-QAM constellation; (b) Block diagram for the conventional QPSK partition algorithm; (c) Block diagram for the modified QPSK partition algorithm.}
\label{fig1}\vspace*{-6pt}
\end{figure}

The partitioning of 16-QAM and the block diagram of QPSK partition algorithm are illustrated in Fig. \ref{fig1}. QPSK partition algorithm distinguishes the QPSK constellation points, whose modulation angles are $\pi/4+n\cdot\pi/2,n\in \{1,2,3,4\}$, from the constellation points on C2 through amplitude decision. The thresholds are commonly determined by Euclidean distance and in the middle of adjacent amplitude levels (the dotted circles in Fig. \ref{fig1}(a)). Then, the modulation phases can be removed and phase noise is estimated by Viterbi-Viterbi algorithm. {\color{red}{The filter with $N+1$ symbols is used to eliminate the effect of ASE noise}}, and phase unwrapping is performed to ensure that the phase estimate conforms to the trajectory of the physical phase.

The block diagram of the proposed jointly optimization algorithm is shown in Fig. \ref{fig1}(c). Since the phase broadening caused by AWGN is smaller at constellation points with larger amplitude, the outer constellation points show more superior performance in CPR \cite{argyris2015high}. However, PS reduces the occurrence of the outer constellation points, which weakens the estimate accuracy of conventional QPSK partition algorithm. Consequently, we increase the filter weight for outer constellation points in PS systems and introduce a parameter $p$ to denote it. Then the phase noise is estimated by
\begin{eqnarray}
\hat{\varphi}_k=\frac{1}{4}\sum_{n=-N/2}^{N/2}((y_{C1})_{k-n}^{4}+p \cdot (y_{C3})_{k-n}^{4}).
\end{eqnarray}
{\color{red}{Where $y_{C1}$ and $y_{C3}$ represent the constellation points of the inner circle and the outer circle after the judgment (the yellow points in Fig. 1(a)). The parameter $p$ is the filter weight introduced by the proposed algorithm, which is used to adjust the weight of the outer constellation points in CPR.}}

\begin{figure}[!t]
\centering
\includegraphics[height= 5.1cm] {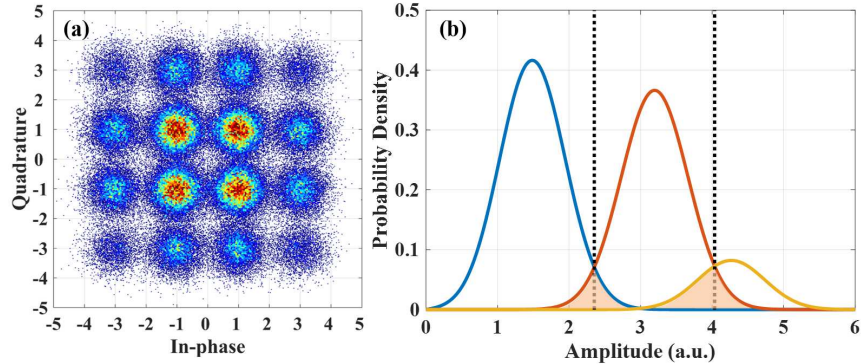}
\caption{(a) PS-16QAM with 3.786 bit/symbol; (b) Probability-weighted PDFs of the individual radii, the dotted lines indicate the optimum amplitude decision thresholds.}
\label{fig2}\vspace*{-6pt}
\end{figure}

In addition to the improvement of filter weight, we modify the decision threshold to get a more precise classification. Conventional amplitude decision designed for uniform distribution is based on Euclidean distance metric, which is sub-optimal for PS constellation. Considering the influence of ASE noise and a-priori probability of the PS signals, the optimal amplitude decision threshold based on MAP is proposed to determine the most likely radius. To understand the calculation of the threshold, we consider the transmitted QAM constellation has $M$ unique radii, and the amplitude of the received symbol, $R$, belonging to one of the radii $A_m$ $(m \in 1,2,\ldots,M)$, is well known to follow the Rician probability density function (PDF):
\begin{eqnarray}
p(R|A_m,\sigma_n^2)=\frac{R}{\sigma_n^2}{\rm exp}(-\frac{R^2+A_m^2}{2\sigma_n^2})I_0(\frac{RA_m}{\sigma_n^2}).
\end{eqnarray}
$I_0(\cdot)$ denotes the zeroth-order modified Bessel function of the first kind. The amplitude PDF of the QAM signal is a mixed distribution, described by a combination of $M$ independent Rician distributions, and weighted according to their a-priori probability $p_m$:
\begin{eqnarray}
p(R|A_1,\ldots,A_m,\sigma_n^2)=\sum_{m=1}^{M} p_m \cdot p(R|A_m,\sigma_n^2).
\end{eqnarray}
Fig. \ref{fig2}(a) shows the PS-16QAM constellation with 3.786 bit/symbol entropy as an example, while SNR is 12 dB. The amplitude PDFs of each radius are plotted in Fig. \ref{fig2}(b). To find the optimal decision threshold, MAP estimation is used to minimize the probability of erroneous decisions. {\color{red}{Error probability can be described as $Pr(A_m|A_{m+1} )+Pr(A_{m+1}|A_m )$. The first term represents the probability that the constellation points are decided as $A_m$ when $A_{m+1}$ is actually transmitted. For 16QAM, the probability of erroneous decision can be expressed as $Pr(A_1|A_2)+Pr(A_2|A_1)+Pr(A_2|A_3)+Pr(A_3|A_2)$, which could be calculated as the area of the shaded part in Fig. \ref{fig2}(b).}} The optimal decision thresholds are represented by the dotted lines.

It should be noted that the proposed MAP-based amplitude decision algorithm requires the knowledge of the a-priori probability and the noise power. In most cases, the a-priori probability of the signal is known to the receiver and noise power could be estimated blindly.

\section{Simulation Setup}
To verify the performance of the proposed algorithm, we implement a Monte-Carlo simulation of PS-16QAM constellations with $2^{17}$ symbols at 56 GBaud. Firstly, a random bit sequence is generated and then fed into the constant composition distribution matcher (CCDM) for probability amplitude shaping (PAS). Shaping factor $\lambda$ is varied from 0 to 0.3 in steps of 0.02 to set Maxwell-Boltzmann distribution. Note that the amplitude shaping of the I and Q branches are independently performed by CCDM and then mapped to QAM signals. The combined laser linewidth is set to 100 kHz. {\color{red}{In order to devote to investigate the impact of PS signals on CPR, we eliminate the interference of other factors and implement a back-to-back transmission.}} AWGN is loaded to vary the SNR within the PS-dominant range, which is 8 dB to 14 dB for PS-16QAM \cite{mello2018interplay}. At the receiver, the conventional QPSK partition algorithm and the proposed MAP-based QPSK partition algorithm with the filter weight optimizing are performed respectively. In order to avoid the impact of cycle slips on estimating the MI, we use a fully supervised way that compares the output of CPR to the transmitted symbols, and compensates for the rotation of $\pi/2$ multiples.

\section{Simulation Results}

\subsection{Optimization of Filter Weight}

\begin{figure}[!t]
\centering
\includegraphics[height= 4.7cm] {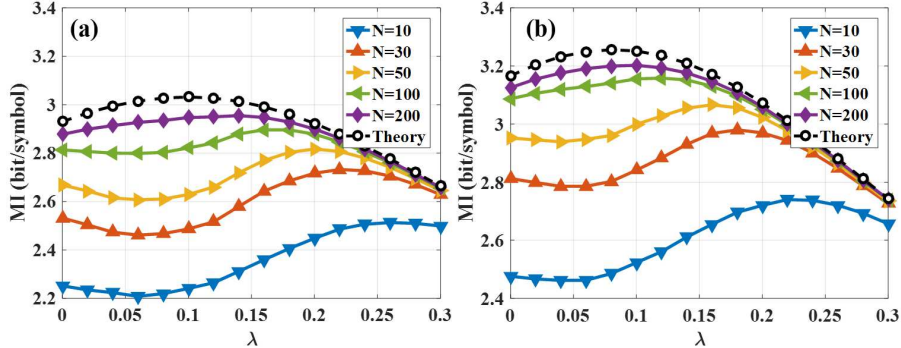}
\caption{MI versus $\lambda$ with conventional QPSK partition algorithm: (a) $\rm{SNR}=9$ dB; (b) $\rm{SNR}=10$ dB.}
\label{fig3}\vspace*{-6pt}
\end{figure}

\begin{figure}[!t]
\centering
\includegraphics[height= 5cm] {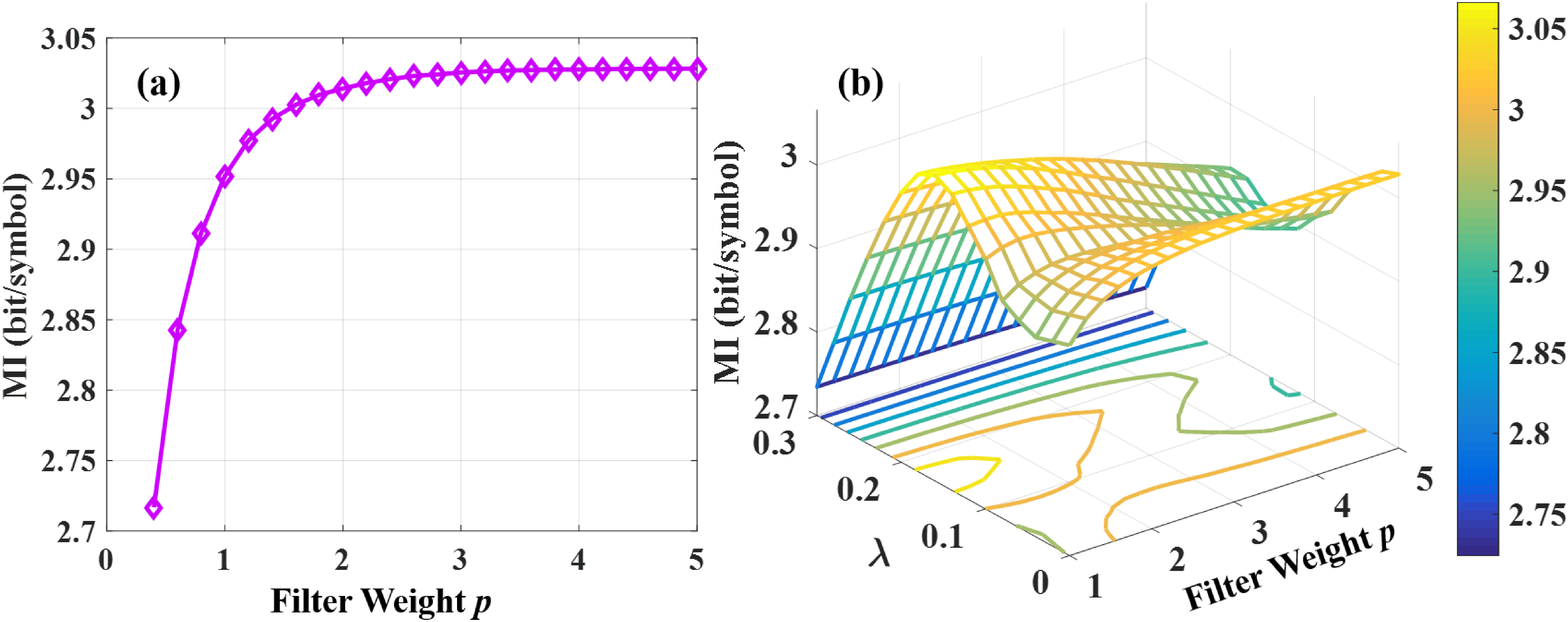}
\caption{MI performance with different filter weight for (a) US-16QAM; (b) PS-16QAM at $\rm{SNR}=10$ dB.}
\label{fig4}\vspace*{-6pt}
\end{figure}

First of all, we investigate the performance of conventional QPSK partition algorithm in PS systems. Fig. \ref{fig3} depicts the MI versus $\lambda$ for different filter length at $\rm{SNR}=9$ dB and $\rm{SNR}=10$ dB. The dashed line indicates the theoretical MI obtained in AWGN channel without phase noise. {\color{red}{It can be observed that the colored curves manifest serious MI impairment with short filters, including two processes of first decline and then recovery. As the shaping factor increases from 0, the channel MI decreases. This is mainly because PS reduces the probability of outer constellation points which show better performance in CPR due to their large Euclidean distance. With the further increase of the shaping factor, the channel MI gradually increases and approaches the theoretical curve. In the second process, the signal power reduction caused by PS becomes obvious gradually. Since the channel SNR is constant, the power of additive noise will also be reduced, contributing to the process of estimation.}} The performance could be restored mainly until $N=200$. It can be understood as a longer filter is required in PS systems to achieve the desired performance exceeding the uniform distribution. As the outer constellation points have a higher tolerance to ASE-induced phase noise and show more superior performance in CPR, the reduction of outer constellation points in PS systems leads to the need for longer filters.

However, considering the computational complexity and feasibility, the filter length is set to 50 in the following. The filter weight could be adjusted to enhance the tolerance of ASE noise, which is limited by filter length. Fig. \ref{fig4}(a) demonstrates the MI as a function of filter weight $p$ for uniform distribution. The uptrend verifies the validity of increasing the filter weight of outer constellation points. Afterwards, filter weight is introduced into probabilistic shaping illustrated in Fig. \ref{fig4}(b). The result reveals that greater weight is advantageous at a smaller shaping factor $(\sim \lambda<0.06)$; on the contrary, when $\lambda$ increases, obviously for $\lambda>0.1$, greater weight leads to more serious impairment. For example, this phenomenon can be clearly observed at $\lambda=0.15$. The maximum MI is 3.07 bit/symbol at $p=1$, however, the MI is degenerated to 2.9 bit/symbol at $p=5$.

\begin{figure}[!t]
\centering
\includegraphics[height= 5.2cm] {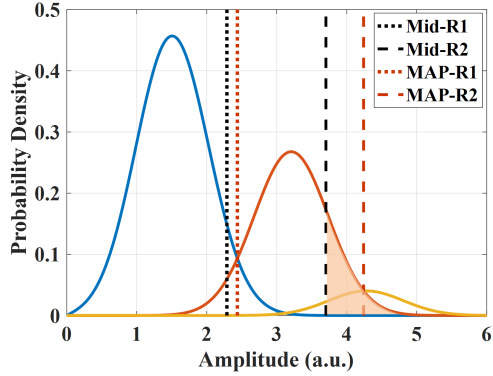}
\caption{The PDFs of the received signals at $\lambda=0.15$, $\rm{SNR}=10$ dB. The black dotted lines indicate conventional median decision thresholds. The red dotted lines indicate the optimized decision thresholds based on MAP.}
\label{fig5}\vspace*{-6pt}
\end{figure}

In order to solve the problem of MI degeneration caused by large filter weight, we investigate the amplitude distribution of received signals. Fig. \ref{fig5} shows the PDFs of the received signals at $\lambda=0.15$, where the performance is sensitive to greater weight in Fig. \ref{fig4}(b). The black dotted lines indicate the conventional median decision threshold $\rm{R1}=(\sqrt{2}+\sqrt{10})/2$ and $\rm{R2}=(\sqrt{10}+3\sqrt{2})/2$. Due to the strong shaping, the probability of outer constellation points is significantly reduced. Affected by AWGN and imperfect decision threshold, increasing the weight of the outer constellation points actually increases the weight of the misjudgments of the middle constellation points more (the shaded part in Fig. \ref{fig5}). The modulation phase of the misjudged points cannot be removed by the fourth power operation, and the performance is damaged consequently. To alleviate the wrong decisions, the thresholds need to be moved to the optimized threshold, which are indicated by the red dotted lines in Fig. \ref{fig5}.

\begin{figure}[!h]
\centering
\includegraphics[height= 5.2cm] {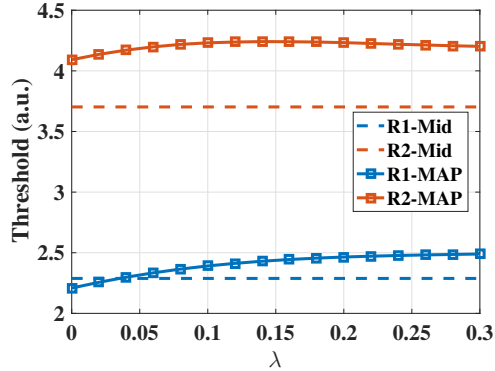}
\caption{Comparison between the conventional median threshold and the MAP-based optimal threshold at $\rm{SNR}=10$ dB.}
\label{fig6}\vspace*{-6pt}
\end{figure}

\subsection{Optimization of Threshold based on MAP}
According to the above analysis, the proposed MAP-based decision threshold could reduce the probability of erroneous decisions effectively. Fig. \ref{fig6} shows the comparison between the conventional median threshold and the MAP-based optimal threshold for different shaping strength at $\rm{SNR}=10$ dB. Affected by AWGN and the non-uniform symbol distribution, the optimized MAP-based threshold has a large deviation from the original threshold, which only considers the Euclidean distance as the criterion. Corresponding to Fig. \ref{fig5}, the decision threshold R2 is moved from $(\sqrt{10}+3\sqrt{2})/2$ to the optimized MAP-based threshold 4.24 showed in Fig. \ref{fig6} when the shaping factor is 0.15. The erroneous decisions are greatly reduced by optimizing the decision threshold.

\begin{figure}[!t]
\centering
\includegraphics[height= 5cm] {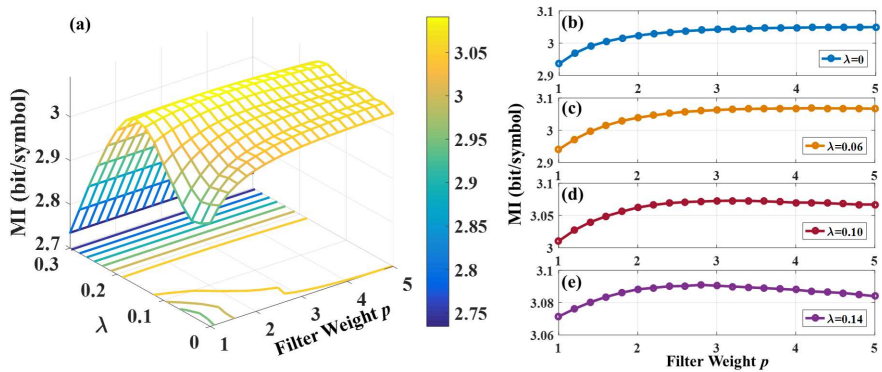}
\caption{(a) MI versus $\lambda$ and $p$ with the proposed MAP-based decision threshold at $\rm{SNR}=10$ dB. (b)-(e) MI versus $p$ for different shaping factors.}
\label{fig7}\vspace*{-6pt}
\end{figure}

The MI performance for different shaping strength and filter weight using the proposed MAP-based algorithm at $\rm{SNR}=10$ dB is shown in Fig. \ref{fig7}. Due to the reduction of misjudgments, the proposed algorithm has a significant improvement at the greater weight compared with the conventional algorithm (see the concave at greater weight in Fig. \ref{fig4}(b)). {\color{red}{The illustrations on the right show the relationship between the MI and the filter weight for different shaping factors. When the shaping factor is less than 0.1, MI increases as the filter weight increases. When the shaping factor is further increased, there will be an impact similar to the Mid-based decision threshold. A large $p$ excessively increases the weight of the misjudged middle circle constellation points, resulting in a slight decrease in performance. However, compared with the Mid-based threshold, the performance degradation under the MAP-based threshold is greatly reduced.}} As an example, the MI increases from 2.9 bit/symbol to 3.08 bit/symbol when $\lambda=0.14$, $p=5$. By contrast, $p=3$ could be selected out as the optimized filter weight with the proposed MAP-based threshold.

\begin{figure}[htp]
\centering
\includegraphics[height= 5.5cm] {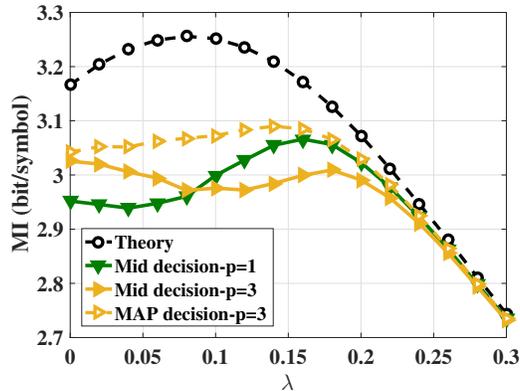}
\caption{MI versus $\lambda$ with the conventional algorithm and the proposed algorithm at $\rm{SNR}=10$ dB. }
\label{fig8}\vspace*{-6pt}
\end{figure}

The comparison of the conventional QPSK partition algorithm and the proposed joint optimization algorithm at $\rm{SNR}=10$ dB is illustrated in Fig. \ref{fig8}. By optimizing the filter weight merely, the MI could be improved when $\lambda<0.08$. As the proportion of the outer constellation points decreasing, greater filter weight makes the error of middle circle being misjudgment enlarged. Therefore, the MI performance is degenerated at a large shaping factor. Subsequently, the MAP estimation is introduced to optimize the decision threshold, which greatly reduces erroneous decisions. Compared with the conventional algorithm, the MI improvement at the theoretical optimal shaping factor $\lambda=0.08$ could attain 0.107 bit/symbol, which is equivalent to a 3.6\% increase in channel capacity.

\begin{figure}[htp]
\centering
\includegraphics[height= 5.2cm] {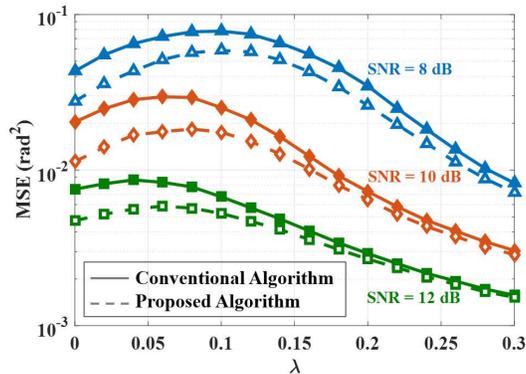}
\caption{Phase noise estimation MSE versus $\lambda$ at $\rm{SNR}=8$ dB, 10 dB and 12 dB, the solid lines indicate the conventional algorithm and the dotted lines indicate the proposed algorithm with the optimized threshold and filter weight $p=3$.}
\label{fig9}\vspace*{-6pt}
\end{figure}

We assess the mean square error (MSE) of the estimated phase noise, where $\rm{MSE}=mean\{(\varphi_k-\hat{\varphi}_k)^2\}$, $\varphi_k$ is the actual phase noise and $\hat{\varphi}_k$ is the estimate value. Fig. \ref{fig9} shows the MSE as a function of $\lambda$ using the conventional algorithm and the proposed MAP-based algorithm with the optimized filter weight $p=3$. The channel SNR is set to 8 dB, 10 dB and 12 dB, respectively. The solid lines represent the conventional algorithm and the dotted lines represent the proposed algorithm. It can be seen that PS aggravates the error in phase noise estimation compared with uniform distribution ($\lambda=0$). The proposed algorithm enhances the tolerance of the ASE-induced phase noise by increasing the filter weight of the outer constellation points, and reduces the misjudgments by optimizing the decision threshold. As a consequence, the estimation error effectively alleviated for each SNR. {\color{red}{The MSE is reduced from $7.6\times10^{-2}$, $3.0\times10^{-2}$ and $8.7\times10^{-3}$ to $5.6\times10^{-2}$, $1.8\times10^{-2}$ and $5.5\times10^{-3}$ at $\rm{SNR}=8$ dB, 10 dB and 12 dB, respectively. The MSE is reduced by nearly half.}}

\begin{figure}[!h]
\centering
\includegraphics[height= 4.7cm] {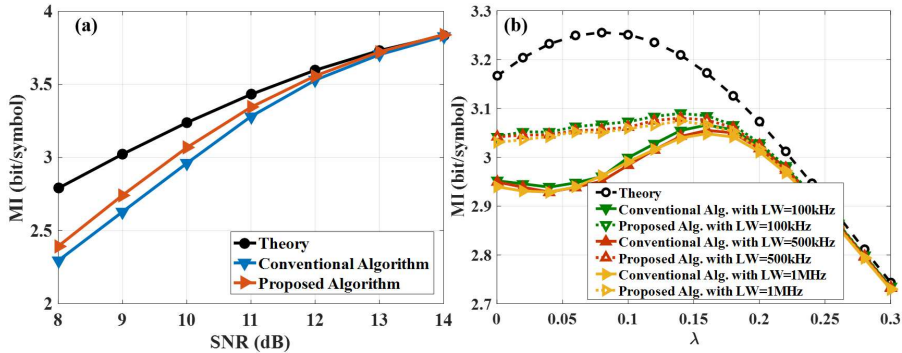}
\caption{(a) Comparison of the conventional algorithm and the proposed algorithm for different SNRs. (b) Comparison of the conventional algorithm and the proposed algorithm for different laser linewidth at $SNR=10$ dB.}
\label{fig10}\vspace*{-6pt}
\end{figure}

The MI performance for the whole SNR interval where PS is dominant at the respective theoretical optimal shaping factors is investigated in Fig. \ref{fig10}(a). The theoretical optimal shaping factors and the optimal decision thresholds for each SNR are summarized in Table \ref{table1}. It can be seen that the proposed joint optimization algorithm is able to partially alleviate the impairment caused by PS and the limitation of filter length. The gap between the theoretical MI and the actually obtained MI could be reduced by 0.1 bit/symbol, 0.114 bit/symbol and 0.107 bit/symbol at a SNR of 8 dB to 10 dB. The channel capacity is increased by 4.2\%, 4.3\% and 3.6\% respectively.

{\color{red}{In order to further investigate the performance of the proposed algorithm, the MI of 100 kHz, 500 kHz and 1 MHz laser linewidth is shown in Fig. \ref{fig10}(b). It can be seen that the gain brought by the proposed algorithm is not affected by the linewidth. At the theoretical shaping factor $\lambda=0.08$, the proposed algorithm can always bring about 0.1 bit/symbol MI gain regardless of the linewidth.}}

\begin{center}
\begin{table}[h]\small
\caption{Summarization of the MAP-based optimal threshold}
\centering
    \begin{threeparttable}
    \setlength{\tabcolsep}{1.3mm}{
    {\begin{tabular}[l]{@{}l|ccccccc}
    \toprule
        SNR(dB) & 8 & 9 & 10 & 11 & 12 & 13 & 14 \\
    \midrule
       Theoretical & \multirow{2}{*}{0.12} & \multirow{2}{*}{0.10} & \multirow{2}{*}{0.08} & \multirow{2}{*}{0.06} & \multirow{2}{*}{0.06} & \multirow{2}{*}{0.04} & \multirow{2}{*}{0.02} \\
       \ optimal $\lambda$ \\[2pt]
    Optimal {\color{red}{R1}} & 2.486 & 2.418 & 2.365 & 2.324 & 2.317 & 2.293 & 2.275 \\[4pt]
    Optimal R2 & 4.551 & 4.368 & 4.218 & 4.096 & 4.015 & 3.937 & 3.875 \\[2pt]
    \bottomrule
    \end{tabular}}}
    \end{threeparttable}
    \label{table1}
\end{table}
\end{center}

\section{Conclusion}

In this paper, the performance of conventional QPSK partition algorithm in probability shaping system was investigated, and a modified QPSK partition algorithm with the MAP-based decision and filter weight optimization was proposed for PS signals. The numerical simulation results show that the performance of conventional QPSK partition algorithm is weakened when PS is implemented. The proposed joint optimization algorithm can reduce the estimation error of phase noise efficiently and relieve the MI impairment. With the proposed algorithm, the channel capacity is increased by 4.2\%, 4.3\% and 3.6\% at a SNR of 8 dB to 10 dB, respectively.

\section*{Acknowledgment}

This work was supported in part by the National Natural Science Foundation of China under Grant 61871044 and Grant 61771062, and in part by the Fund of the State Key Laboratory of IPOC (BUPT) (No. IPOC2018ZT08), P. R. China.

\section*{References}

\bibliography{reference}

\end{document}